\documentclass[final,5p,times,twocolumn]{elsarticle}
\usepackage{graphicx}
\graphicspath{{.}{./EPS/}}
\usepackage{amssymb}
\usepackage{amsmath}
\usepackage{amsfonts}
\DeclareMathOperator{\signum}{sgn}
\ifx\openone\undefined
 \newcommand{\openone}{\leavevmode\hbox{\small1\kern-3.8pt\normalsize1}}
\fi

\usepackage{bm}
\usepackage{array}
\newcolumntype {s}[1]{@{\hspace{#1}}} % space

\newcommand* {\vek}[1]{{{\ensuremath{\bm{\mathrm{#1}}}}}}

\newcommand* {\kk}{\vek{k}}

\newcommand* {\ee}{\ensuremath{\mathrm{e}}}

\newcommand* {\Ts}{\textstyle}

\newcommand* {\frack}[2]{{\Ts\frac{#1}{#2}}}

\newcommand* {\bra}[1]{\ensuremath{\langle {#1} |}}
\newcommand* {\ket}[1]{\ensuremath{| {#1} \rangle}}
\newcommand* {\bracket}[2]{\ensuremath{\langle {#1} | {#2} \rangle}}

\newcommand* {\phik}[1][]{\varphi_{#1\vek{k}}}
\newcommand* {\typF}{{\ensuremath{\mathrm{f}}}}
\newcommand* {\typB}{{\ensuremath{\mathrm{b}}}}
\newcommand{\marr}[2][c]{\left({\renewcommand{\arraystretch}{0.8}%
  \tabcolsep 0pt\begin{array}{@{}#1@{}} #2 \end{array}}\right)}

\biboptions{sort&compress}
\journal{Physics Letters A}

\begin{document}
\begin{frontmatter}
\title{Time reversal of pseudo-spin $1/2$ degrees of freedom}
\author[ifs,anl,niu]{R. Winkler}
\author[ifs,ctcp]{U. Z\"ulicke\corref{contact}}
\ead{u.zuelicke@massey.ac.nz}

\address[ifs]{Institute of Fundamental Sciences and MacDarmid Institute for Advanced
Materials and Nanotechnology, Massey University, Manawatu Campus, Private Bag
11~222, Palmerston North 4442, New Zealand}
\address[anl]{Materials Science Division, Argonne National Laboratory, Argonne, IL
60439, USA} 
\address[niu]{Department of Physics, Northern Illinois University, DeKalb, IL 60115,
USA}
\address[ctcp]{Centre for Theoretical Chemistry and Physics, Massey University,
Albany Campus, Private Bag 102~904, North Shore MSC, Auckland 0745, New
Zealand}

\cortext[contact]{Corresponding author}

\begin{abstract}
  We show that pseudo-spin $1/2$ degrees of freedom can be
  categorized in two types according to their behavior under time
  reversal. One type exhibits the properties of ordinary spin whose
  three Cartesian components are all odd under time reversal. For
  the second type, only one of the components is odd while the other
  two are even. We discuss several physical examples for this
  second type of pseudo-spin and highlight
  observable consequences that can be used to distinguish it from
  ordinary spin.
\end{abstract}

\begin{keyword}
%% keywords here, in the form: keyword \sep keyword
  foundations of quantum mechanics \sep reversal of motion \sep
  dynamical symmetry

%% MSC codes here, in the form: \MSC code \sep code
%% or \MSC[2008] code \sep code (2000 is the default)

\end{keyword}

\end{frontmatter}

%%
%% Start line numbering here if you want
%%
% \linenumbers

\section{Introduction}

The behavior under time reversal (TR; also denoted \emph{reversal of
 motion\/}) is one of the most fundamental characteristics of a
quantum system~\cite{wig32, sac87}. It determines, e.g., level
degeneracies~\cite{sak94} and the statistics of energy-level
spacings~\cite{bro81} in closed systems, electric transport in
phase-coherent quantum circuits~\cite{bee97,akk07}, and the possible
channels for pairing of electrons to form a superconducting
Cooper-pair condensate~\cite{tin96}. Formally, the TR operation can
be represented by an anti-unitary operator $\hat\theta$ that is,
however, specific to the particular choice of base kets~\cite{sac87,
 sak94}. Quite generally, we may write
\begin{equation}
\label{eq:general}
\hat\theta = \hat U \, {\mathcal C} \quad ,
\end{equation}
where $\hat U$ denotes a unitary operator and ${\mathcal C}$ is
complex conjugation. For simple quantum systems, the explicit form
of the TR operator $\hat\theta$ is well known~\cite{sac87, sak94}.
Recent efforts were aimed at generalizing the TR operation to more
complex systems, e.g., those having internal degrees of
freedom~\cite{sud95}.

We focus here on TR of quantum systems that carry an effective SU(2)
(pseudo or real) spin degree of freedom that may be half-integer or
integer. We show that SU(2) symmetry allows for two fundamentally
different behaviors under TR. As TR is an anti\-unitary symmetry
independent of the unitary symmetry elements in SU(2), it needs to
be determined based on physical considerations which TR behavior
applies to a particular system. This result has important
consequences for effective pseudo-spin descriptions that are widely
utilized. Classic examples include Schwinger's oscillator model of
angular momentum~\cite{sch65} (see also Sec.~3.8 in
Ref.~\cite{sak94}), nuclear isospin~\cite{gre94}, and the ammonia
molecule~\cite{fey65}. More recently, the pseudo-spin concept has
been ubiquitous in the context of quantum information
processing~\cite{nie00}. Other pseudo-spin-carrying entities of
current interest include the massless Dirac-electron-like
quasiparticles in graphene~\cite{cas09} and the persistent spin
helix in quasi-twodimensional semiconductor systems with fine-tuned
spin-orbit couplings~\cite{ber06,kor09}. We will discuss these
particular examples and elucidate experimentally observable
ramifications for the two different types of pseudo-spins.

\section{Time reversal of a pseudo-spin: General properties}

We start by considering the textbook example of an SU(2)
angular-momentum algebra involving the three operators $\hat J_j$
with $j=1,2,3$, satisfying the commutation relations
\begin{equation}
  \label{eq:commutator}
  \big[ \hat J_j\, , \, \hat J_k \big] = i \, \epsilon_{jkl}\, \hat J_l \quad .
\end{equation}
In general, kinematically relevant physical quantities are either
even or odd under TR~\cite{note:evenodd}. Allowing for both
possibilities for each operator $\hat J_j$, we write $\hat\theta \,
\hat J_j \, \hat \theta^{-1} = \xi_j \, \hat J_j$, with $\xi_j = \pm
1$. As the commutators (\ref{eq:commutator}) need to be preserved
under TR, the three coefficients $\xi_j$ cannot be independent.
Rather, they must satisfy the condition $\xi_1 \xi_2 = - \xi_3$,
which restricts the possible TR behavior to two cases:
(\typF)~$\xi_1=\xi_2=\xi_3 \equiv - 1$, or (\typB)~$\xi_1 = -\xi_2 =
-\xi_3 \equiv - 1$, with permutations of indices allowed.
Case~(\typF) implies that all operators $\hat J_j$ are odd under TR,
which is the behavior found for the Cartesian components of orbital
and ordinary-spin angular momentum \cite{sak94}. In case~(\typB),
\emph{only one\/} of the operators $\hat J_j$ is odd under TR
(without loss of generality chosen here to be $\hat J_1$), and the
other two ($\hat J_2$ and $\hat J_3$) are both \emph{even\/}. For an
SU(2) invariant system of type (\typB), one of the Cartesian
components of $\hat{\vek{J}}$ is thus always distinguished by its
behavior under TR.

The TR behavior associated with cases (\typF) and (\typB)
generally leads to qualitatively different physical properties. For
half-integer (pseudo-) spin systems, a basic feature distinguishing
the two cases is given by the fact that $\hat\theta_\typF^2 =
-\openone$, whereas case (\typB) implies $\hat\theta_\typB^2 =
+\openone$~\cite{note:fermibose}. To prove these relations, we
construct the TR operator for cases (\typF) and (\typB) in the usual
representations \cite{sak94}, where the matrix elements of $\hat
J_1$ and $\hat J_3$ are real, while those for $\hat J_2$ are
imaginary. Given the general relation expressed in
Eq.~(\ref{eq:general}), it is only necessary to find unitary
transformations $\hat U_\typF$ and $\hat U_\typB$ (i.e., rotations
in spin space) that yield the required transformation properties of
the operators $\hat J_j$. As $\hat J_2$ is odd under complex
conjugation while $\hat J_1$ and $\hat J_3$ are even, it is
straightforward to find $\hat U_\typF = \exp\big( i \,\pi \hat J_2
\big)$ and $\hat U_\typB = \exp\big( i \,\pi \hat J_3 \big)$. Noting
that ${\mathcal C}\, \hat U_\typF \, {\mathcal C} = \hat U_\typF$,
we find $\hat\theta_\typF^2 = \hat U_\typF^2 \equiv \exp \big(2i \,
\pi \hat J_2 \big)$ amounts to a $2\pi$ rotation in spin space,
which yields $\hat\theta_\typF^2 = -\openone$ for half-integer spin.
Conversely, ${\mathcal C}\, \hat U_\typB \, {\mathcal C} = \hat
U_\typB^{-1}$ and thus $\hat\theta_\typB^2 = \hat U_\typB \, \hat
U_\typB^{-1} \equiv +\openone$.

The sign of $\hat\theta^2$ has direct experimental consequences for
the quantum interference of amplitudes for TR-related physical
scenarios. Such interference plays a role, e.g., in particle
interferometers~\cite{rau75,wer75} and for the phase-coherent propagation
of electrons and light through disordered media~\cite{akk07}.

\section{Physical realizations of case-(\typB) pseudo-spin}

Whether a (pseudo-) spin degree of freedom belongs to case (\typF)
or (\typB) depends on the details of the particular system under
consideration. It is well-known~\cite{sak94} that orbital and
ordinary-spin angular momentum belong to case~(\typF). In contrast,
the possibility to have the TR properties of case~(\typB) has been
under-appreciated, even though physical realizations exist. For
example, it is well-known that the third Cartesian component of
nuclear isospin must be even under TR by virtue of electric-charge
conservation, and some implications have been considered in previous
work~\cite{sac87, sud95}. Below we further elucidate the general TR
properties of type-(\typB) pseudo-spins using isospin as an example,
and also Schwinger's bosonic model of angular momentum~\cite{sch65},
which is equivalent to the isotropic two-dimensional (2D) harmonic
oscillator~\cite{jau40, gol00a}. Finally, we show that case~(\typB)
applies also to the persistent spin helix~\cite{ber06}.

\subsection{Nuclear isobaric spin}

The isospin-1/2 model is used to describe states of the
nucleon~\cite{gre94}, with eigenstates of $\hat J_3$ being
associated with the proton and neutron, respectively. The
electromagnetic interaction distinguishes between the two states,
and conservation of electric charge requires that all relevant
states are eigenstates of $\hat J_3$. In addition, charge
conservation mandates the invariance of $\hat J_3$ under TR. In the
usual representation where $\hat J_3$ is real, this implies
$\hat\theta\, \hat J_3 \, \hat\theta^{-1} = \hat U\, \hat J_3 \,
\hat U^{-1} = \hat J_3$, i.e., $\hat U$ must be a rotation around
the 3-axis in isospin space. As the physically relevant isospin
states are eigenstates of $\hat J_3$, $\hat U$ was not specified
further in previous work~\cite{sac87}. However, it is
straightforward to verify that $\hat\theta^2 = \hat U \, \hat U^{-1}
= +\openone$, i.e., isospin is an example for the type-(\typB)
pseudo-spin.

\subsection{Isotropic 2D harmonic oscillator and Schwinger model}

It was realized early on~\cite{jau40} that the three dynamic
invariants of the isotropic 2D harmonic oscillator~\cite{gol00a}
correspond to a dynamical SU(2) symmetry. Using dimensionless
coordinate and momentum variables, in which the Hamiltonian reads
$H=(p_1^2+q_1^2)/2 + (p_2^2+q_2^2)/2$, the three conserved
quantities can be expressed as $\hat J_1 = (q_1 p_2 - q_2 p_1)/2$,
$\hat J_2 = (q_1 q_2 + p_1 p_2)/2$, and $\hat J_3 = [p_2^2 + q_2^2 -
(p_1^2 + q_1^2)]/4$. They correspond to the orbital angular momentum
of the oscillator ($\hat J_1$), the energy difference for motions in
the two perpendicular in-plane directions ($\hat J_3$), and the
phase difference between oscillations in those directions ($\hat
J_2$). Straightforward calculation based on the canonical
commutation relations $[ q_j\, , p_k ] = i \delta_{jk}$ and $[ q_i\,
, q_j ] = [ p_i\, , p_j ] = 0$ establishes that the $\hat J_j$
satisfy Eq.~(\ref{eq:commutator}). Furthermore, the equivalence
between this system and Schwinger's bosonic model of angular
momentum~\cite{sch65,sak94} becomes apparent when the quantities
$\hat J_j$ are expressed in terms of creation and annihilation
operators $a_k^\dagger=(q_k - i p_k)/\sqrt{2}$, $a_k=(q_k + i
p_k)/\sqrt{2}$ for the two 1D oscillators.

As the coordinate (momentum) components are even (odd) under TR, it
follows that $\hat J_1$ is odd but both $\hat J_2$ and $\hat J_3$
are even so that this system is a realization of type (\typB). Even
when the Schwinger-model description is applied to a general
two-level system and, thus, the underlying oscillator degree of
freedom is abstract, the definition of $\hat J_3 \equiv (a_2^\dagger
a_2 - a_1^\dagger a_1)/2$ in terms of the occupation numbers of the
two levels implies that $\hat J_3$ must be even under TR. Hence,
many pseudo-spin models used in condensed-matter
physics~\cite{lie04} and quantum optics~\cite{wal08} belong to type
(\typB).

\subsection{Persistent spin helix}

The persistent spin helix is a recently
discovered~\cite{ber06,kor09} collective excitation present in
certain quasi-2D semiconductor systems with fine-tuned spin-orbit
couplings. In terms of second-quantized operators $c_{\vek{k},
 \uparrow (\downarrow)}^\dagger$ and $c_{\vek{k}, \uparrow
 (\downarrow)}$ that, respectively, create and annihilate an
electron with wave vector $\vek{k}$ and spin-up (spin-down), the
following operators associated with the persistent spin helix are
defined
\begin{subequations}
\begin{eqnarray}
S_{\vek{Q}}^+ &=& \sum_\vek{k} c^\dagger_{\vek{k}+\vek{Q}, \uparrow} \,
c_{\vek{k}, \downarrow} \,\, , \qquad S_{\vek{Q}}^-
= \sum_\vek{k} c^\dagger_{\vek{k},
\downarrow} \, c_{\vek{k}+\vek{Q},\uparrow} \,\,\, , \\
S_z &=& \frac{1}{2} \sum_{\vek{k}} \left( c^\dagger_{\vek{k}, \uparrow} \,
  c_{\vek{k},
\uparrow} - c^\dagger_{\vek{k}, \downarrow} \, c_{\vek{k}, \downarrow}
\right) \quad ,
\end{eqnarray}
\end{subequations}
where $\vek{Q}$ is a function of the spin-orbit coupling strength in
the system \cite{ber06}. If we define $S_x = (S_{\vek{Q}}^+ +
S_{\vek{Q}}^- )/2$ and $S_y = (S_{\vek{Q}}^+ - S_{\vek{Q}}^-
)/(2i)$, it can be shown that the components $S_j$ obey Eq.\
(\ref{eq:commutator}).
Using the fact that both $\vek{k}$ and spin $\uparrow,\downarrow$
are odd under TR, we find immediately $\hat\theta \, S_z \,
\hat\theta^{-1} = - S_z$. Similarly, we get $\hat\theta \,
S_\vek{Q}^\pm \,\hat\theta^{-1} = S_\vek{Q}^\mp$ which implies that
$S_x$ and $S_y$ are even under TR. Hence, the SU(2) degree of
freedom associated with the persistent spin helix is a type-(\typB)
pseudo-spin.

\section{Conventional and unconventional TR of 2D massless Dirac
 particles}
\label{sec:DiracTR}

\begin{figure}[tbp]
  \includegraphics[width=1.0\columnwidth]{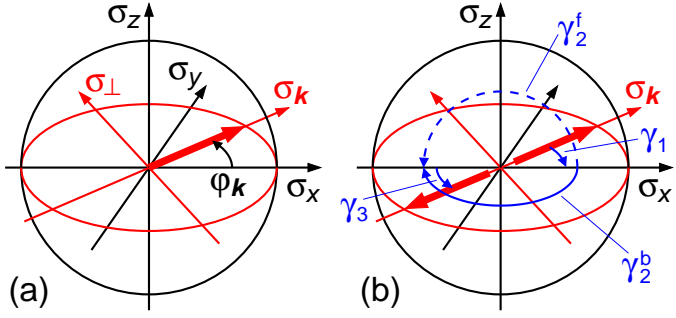}
  \caption{\label{fig:IVTR} (a) Bloch sphere for the pseudospin of
   2D massless Dirac particles. The
   pseudo-spin for wave vector $\vek{k}$ is characterized by a
   rotated coordinate system $(\sigma_\vek{k}, \sigma_\perp,
   \sigma_z)$ with $\sigma_\vek{k}$ parallel to $\vek{k}$. (b)~Time
   reversal (TR) of pseudospin parallel to $\vek{k}$ can be related
   to TR of pseudospin parallel to the $x$ axis via two rotations
   $\gamma_1$ and $\gamma_3$ around the $z$ axis by angles $-\phik$
   and $\phik$. Depending on whether TR of pseudospin parallel to
   the $x$ axis is achieved via a $\pi$ rotation around the $y$ axis
   ($\gamma_2^\typF$) or around the $z$ axis ($\gamma_2^\typB$), one
   obtains $\hat \vartheta_\typF$ or $\hat \vartheta_\typB$.}
\end{figure}

In the previous section we discussed examples, where the components
$\hat{J}_j$ of the pseudo-spin were conserved, $[\hat{H}, \hat{J}_j]
= 0$ so that the TR properties of $\hat{J}_j$ could be discussed
separately from the orbital motion. As a classic example where the
(pseudo-) spin degree of freedom is coupled to the orbital motion,
we now discuss particles confined to the $xy$ plane that are
described by a massless Dirac Hamiltonian. Using a plane-wave
($\vek{k}$) basis, the Hamiltonian becomes $\hat H(\vek{k}) =\hbar v
k \sigma_\vek{k}$, with velocity $v$ and the (pseudo-) spin
operator~$\sigma_\vek{k}$
\begin{equation}
\sigma_\vek{k} = \sigma_x \cos\phik + \sigma_y \sin\phik
\equiv \marr[cc]{0 & \ee^{-i\phik} \\ \ee^{i\phik} & 0} \, .
\end{equation}
Here the $\sigma_j$ denote the familiar Pauli matrices, and $\phik$
is the angle between $\vek{k}$ and the $k_x$ axis [see
Fig.~\ref{fig:IVTR}(a)]. The eigenvalues of $H (\vek{k})$ are
$\pm\hbar v k$, and the corresponding eigenstates are
\begin{equation}
\label{eq:dirac_eigen}
\ket{\vek{k},\pm} \equiv \ee^{i\vek{k}\cdot\vek{r}} \ket{\pm}_\vek{k}
\equiv  \frac{\ee^{i\vek{k}\cdot\vek{r}}}{\sqrt{2}}\,
\marr{\ee^{-i\phik/2} \\ \pm \ee^{i\phik/2}}.
\end{equation}
It is reasonable to enable a unique identification of the
eigenstates by restricting $-\pi < \phik \le \pi$.

Reversal of motion should map the eigenstates of $\hat
H (\vek{k})$ as follows:
\begin{equation}
  \hat \vartheta \ket{\vek{k},\pm} 
  = \eta(\kk) \ket{-\vek{k},\pm} \quad ,
\end{equation}
where $\eta(\vek{k})$ stands for an arbitrary phase, i.e., $\hat
\vartheta$ should reverse the wave vector while preserving the
energy. Inspection of Eq.~(\ref{eq:dirac_eigen}) yields the relation
$\ket{\pm}_{-\vek{k}} = i \signum(\phik)\, \ket{\mp}_\vek{k}$,
consistent with $\sigma_{-\vek{k}} = -\sigma_\vek{k}$. Thus
$\sigma_\vek{k}$ needs to be odd under the TR operation, to ensure
TR invariance of $\hat H (\vek{k})$. For a particle confined to two
spatial dimensions, this condition does not uniquely specify the TR
operation. Rather, two possible scenarios exist for how the
direction of $\ket{\vek{k},\pm}$ can be reversed for any given
$\vek{k}$. This is illustrated in Fig.~\ref{fig:IVTR}(b).
Mathematically, the two different TR operations are given by
\begin{subequations}
\begin{eqnarray}\label{eq:AndoTR}
\hat\vartheta_\typF &=& \exp (-\frack{i}{2} \phik \, \sigma_z) \,
\exp (\frack{i}{2} \pi \, \sigma_y) \: {\mathcal C} \, \exp (\frack{i}{2}
\phik\,\sigma_z) \nonumber \\ &=& \exp (\frack{i}{2} \pi \, \sigma_y)
\, {\mathcal C} \equiv i \sigma_y \; {\mathcal C} \quad , \\ \label{eq:ourSimp}
\hat\vartheta_\typB &=& \exp (-\frack{i}{2} \phik \, \sigma_z) \,
\exp [\frack{i}{2} \signum (\phik) \pi \, \sigma_z] \: {\mathcal C}
\, \exp (\frack{i}{2} \phik\,\sigma_z) \nonumber \\ &=&
\exp (-\frack{i}{2} [\phik[-] + \phik] \, \sigma_z) \; {\mathcal C} \quad .
\end{eqnarray}
\end{subequations}

It is straightforward to verify that $\hat H (\vek{k})$ is invariant
under both types of TR transformation. However,
$\hat\vartheta_\typF$ and $\hat\vartheta_\typB$ can be distinguished
by the way the pseudo-spin operators $\sigma_z$ and $\sigma_\perp$
are transformed. [$\sigma_\perp$ is the pseudo-spin in-plane
component perpendicular to $\sigma_\vek{k}$, see
Fig.~\ref{fig:IVTR}(a).] More specifically, both $\sigma_z$ and
$\sigma_\perp$ turn out to be odd under $\hat\vartheta_\typF$,
whereas they are even under $\hat\vartheta_\typB$. Hence, which of
the two operators $\hat\vartheta_{\typB,\typF}$ constitutes the
proper TR operation for a particular system will depend on whether
the actual physical quantities represented by $\sigma_z$ and
$\sigma_\perp$ are even or odd under TR.

\begin{figure}[tbp]
  \includegraphics[width=1.0\columnwidth]{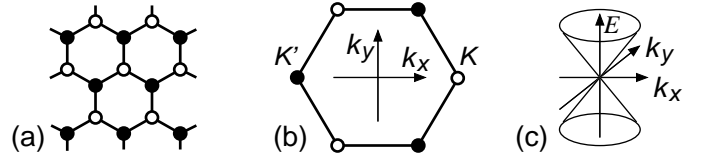}
  \caption{\label{fig:lattice} (a) Honeycomb lattice of graphene.
  Atoms in sublattice $A$ ($B$) are marked with open (closed)
  circles. (b) Brillouin zone and its two inequivalent corner points
  $\vek{K}$ and $\vek{K}'$. The remaining corners are related with
  $\vek{K}$ or $\vek{K}'$ by reciprocal lattice vectors. (c)
  Dispersion $E(k)$ near the $\vek{K}$ point.}
\end{figure}
Before closing this Section, we briefly discuss time-reversal
properties of the sublattice-related pseudo-spin degree of freedom
carried by electronic quasiparticles in graphene~\cite{cas09}. This
material consists of a sheet of carbon atoms arranged on a honey\-comb
lattice, with two inequivalent sublattices labeled $A$ and $B$
[Fig.~\ref{fig:lattice}(a)]. Its Brillouin zone is also hexagonal,
with two inequivalent corner points $\vek{K}$ and $\vek{K'}=
-\vek{K}$ [Fig.~\ref{fig:lattice}(b)]. Most remarkably, the energy
dispersion near $\vek{K}$ and $\vek{K'}$ is linear, and the Fermi
energy of undoped graphene is exactly at the crossing point
[Fig.~\ref{fig:lattice}(c)]. Hence, the low-energy electronic
excitations in graphene are considered to be analogs of 2D massless
Dirac electrons~\cite{cas09, sem84, div84, hal88}.

In situations where Umklapp processes are absent, it is tempting to
treat the 2D Dirac cones at the $\vek{K}$ and $\vek{K}^\prime$
valleys separately and also define an effective intra-valley
time-reversal operation~\cite{and98,suz02}. As the pseudo-spin
operator $\sigma_z$ represents the quasiparticle location on
sublattice $A$ or $B$, it must be even under time reversal (as
expected for a real-space position operator). $\vartheta_\typB$ of
Eq.~(\ref{eq:ourSimp}) exhibits the required property to leave
$\sigma_z$ invariant (in contrast to $\vartheta_\typF$), suggesting
it to be a proper intra-valley time-reversal operator. However, it
should be noted that the ordinary TR for Bloch electrons in graphene
is given by $\hat\theta ={\mathcal C}$ which couples the two
valleys~\cite{suz02}. In order to apply the discussion of this
Section to graphene, a restriction of anti-unitary operators to
individual valleys needs to be formulated.

\section{Observable signatures of the two types of pseudo-spin}

Whether a particular realisation of pseudo-spin belongs to type
(\typF) or (\typB) is not a purely academic question, as their
different TR properties affect physical observables. For example, in
the case of 2D Dirac particles considered in the previous Section, a
(possibly random) potential $V(\vek{r})\sigma_z$ does \emph{not\/}
break $\hat\vartheta_\typB$-TR symmetry, whereas such a potential
will break the $\hat\vartheta_\typF$-TR symmetry~\cite{ber87}. As a
result, the level statistics of chaotic quantum dots formed by a
non-integrable mass-confinement of 2D Dirac particles will be
different in the two cases~\cite{ber87}. While this example is
specific to the case of 2D Dirac particles, we will now discuss a
more general observable difference exhibited by half-integer
pseudo-spin particles of type-(\typF) and (\typB) that are scattered
by a spin-independent random potential but are free otherwise. The
quantum interference of TR-related backscattering amplitudes turns
out to be different for the two pseudo-spin types and can thus serve
to distinguish them experimentally.

We denote scattering states of the particle by $\ket{\vek{k},s}$,
where $\vek{k}$ denotes the wave vector and $s$ is the pseudo-spin
quantum number w.r.t.\ some specific basis. The probability
amplitude for a particular backscattering process involving $n$
scattering events due to a disorder potential is then proportional
to
\begin{eqnarray}
{\mathcal A}_{\vek{k}_1, s_1, \dots, \vek{k}_n, s_n} &=&
\bracket{-\vek{k},s}{\vek{k}_n,s_n}
\bracket{\vek{k}_n,s_n}{\vek{k}_{n-1},s_{n-1}} \dots \nonumber \\ &&
\dots \bracket{\vek{k}_2,s_2}{\vek{k}_1,s_1} 
\bracket{\vek{k}_1,s_1}{\vek{k},s} \, .
\end{eqnarray}
In the typical situation where disorder potentials are invariant
under TR, each such process has a ``partner'' process
\begin{eqnarray}
\tilde{\mathcal A}_{\vek{k}_1, s_1,\dots, \vek{k}_n, s_n} &=&
\bracket{-\vek{k},s}{\hat\theta(\vek{k}_1,s_1)}
\bracket{\hat\theta(\vek{k}_1,s_1)} {\hat\theta(\vek{k}_2,s_2)} \dots
\nonumber \\ && \hspace{-1cm} \dots
\bracket{\hat\theta(\vek{k}_{n-1},s_{n-1})} {\hat\theta(\vek{k}_n,s_n)}
\bracket{\hat\theta(\vek{k}_n,s_n)}{\vek{k},s} \, ,
\end{eqnarray}
where scattering occurs in time-reversed order. Using the
relations~\cite{sak94} $\bracket{a}{b} = \bracket{\hat\theta
 b}{\hat\theta a}$ and $\ket{\hat\theta(\vek{k},s)} =
\ket{-\vek{k},s}$, we find
\begin{eqnarray}
\tilde{\mathcal A}_{\vek{k}_1, s_1 \dots, \vek{k}_n, s_n} &=&
\bracket{-\vek{k}, s}{\hat\theta^2(\vek{k}_n,s_n)}
\bra{\hat\theta^2(\vek{k}_n, s_n)} \dots 
\nonumber \\ && \dots
\ket{\hat\theta^2(\vek{k}_1,s_1)}
\bracket{\hat\theta^2(\vek{k}_1,s_1)} {\hat\theta^2 (\vek{k},s)}
\nonumber \\
&=& \signum (\hat\theta^2) \,\, {\mathcal A}_{\vek{k}_1, s_1\dots,
\vek{k}_n, s_n} \quad .
\end{eqnarray}

The total probability for backscattering of a particle is the
modulus square of the sum over the probability amplitudes of all
possible back-scattering processes. If the sign of $\hat\theta^2$ is
positive (negative), scattering processes related by TR will
interfere constructively (destructively), leading to reduced
(enhanced) particle transmission through the medium. Thus we suggest
that quantum-coherent transport provides an avenue for
distinguishing between half-integer type-(\typF) and type-(\typB)
pseudo-spins, because $\hat\theta_{\typF}^2 = -\hat\theta_{\typB}^2
\equiv -1$. Direct observation of the backscattering probability
would facilitate such an experimental distinction.

\section{Conclusions}

We have shown that (pseudo-) spin degrees of freedom can be classed
into two types according to their time-reversal properties. For type
(\typF), which includes ordinary spin angular momentum, all three
spin components are odd under time reversal. For type (\typB), only
one of the components is odd under time reversal while the remaining
two are even. This case includes nuclear isospin and the dynamic
SU(2) symmetry of the 2D isotropic harmonic oscillator.

As an example of a system for which (pseudo-) spin and orbital
motion are coupled, we discussed time reversal of massless Dirac
particles confined to move in two spatial dimensions. It was found
that the dynamics of the system again allows for time-reversal
operations of the two types (\typF) and (\typB). Only by considering
the properties of additional observables represented, e.g., by the
$z$ component of pseudo-spin is it possible to uniquely determine
the form of the time-reversal operation.

Which type of time-reversal operation is realised in a particular
system has measurable consequences. For example, whether itinerant
particles with half-integer (pseudo-) spin belong to type (\typF) or
(\typB) manifests itself in an experimentally observable way by the
relative sign of probability amplitudes associated with
time-reversal-related back-scattering scenarios. This and several
more observable ramifications associated with randomness/chaotic
dynamics~\cite{bee97} arise because the squared anti-unitary
operators associated with the time-reversal transformation in the
two cases differ by their sign.

\section*{Acknowledgements}

This work is supported by the Marsden Fund Council (contract
MAU0702) from Government funding, administered by the Royal Society
of New Zealand. We thank the Kavli Institute for Theoretical Physics
China at the Chinese Academy of Sciences for hospitality and support
during the final stages of writing this article. Work at Argonne was
supported by DOE BES under Contract No.\ DE-AC02-06CH11357.
Discussions with M.~Fortner, M.~J\"a\"askel\"ainen, A.H.~MacDonald,
and A.I.~Signal are gratefully acknowledged.

%\bibliographystyle{model1a-num-names}
%\bibliography{general,mesophys,graphene,Roland_adjust}
%\bibliography{journal,english,PLApaper,semi,graphene,quantum,buch,my}

\begin{thebibliography}{28}
\bibitem{note:evenodd}
\bibinfo{note}{In the most general scenario, we can decompose every
observable into \emph{components\/} that are even or odd under TR.}
\bibitem{note:fermibose}
\bibinfo{note}{The property $\hat\theta_\typF^2 = -\openone$ is closely
related to the fermionic nature of primitive particles carrying half-integer
spin of type (\typF). Conversely, $\hat\theta_\typB^2 = +\openone$ for
case (\typB) implies bosonic behavior for such primitive particles. In reality,
pseudo-spin carrying particles have additional degrees of freedom and
thus can be bosonic or fermionic. In our discussions presented here, we
focus only on single-particle phenomena and particle statistics does not matter.}
\end{thebibliography}

\end{document}